\documentclass{elsarticle}

\usepackage{epsfig}
\usepackage{a4wide}
\usepackage{amsmath}

\begin{document}

\vspace{\baselineskip}

\title{A novel subtraction scheme for double-real radiation at NNLO}

    \author[aachen]{M. Czakon}

    \address[aachen]{
      Institut f\"ur Theoretische Teilchenphysik und Kosmologie,
      RWTH Aachen University,\\  D-52056 Aachen, Germany
    }

\cortext[thanks]{Preprint number: TTK-10-31}

\begin{abstract}

  \noindent
  A general subtraction scheme, {\tt STRIPPER} (SecToR Improved
  Phase sPacE for real Radiation), is derived for the evaluation of
  next-to-next-to-leading order (NNLO) QCD contributions from
  double-real radiation to processes with at least two particles in
  the final state at leading order. The result is a Laurent expansion
  in the parameter of dimensional regularization, the coefficients of
  which should be evaluated by numerical Monte Carlo integration. The
  two main ideas are a two-level decomposition of the phase space,
  the second one factorizing the singular limits of amplitudes, and a
  suitable parameterization of the kinematics allowing for derivation
  of subtraction and integrated subtraction terms from eikonal factors
  and splitting functions without non-trivial analytic integration.

\end{abstract}

\maketitle

%%%%%%%%%%%%%%%%%%%%%%%%%%%%%%%%%%%%%%%%%%%%%%%%%%%%%%%%%%%%%%%%%%%%%%%%%%%%%%%%

\section{Introduction}

Compared to the number of phenomenological applications, where NNLO
QCD corrections are indispensable, the effort put into the
construction of general subtraction schemes for real radiation at this
level of perturbation theory is astounding. The main problems
encountered are either, that the method is not general and requires
tedious adaptation to every specific problem, or that there are many
highly non-trivial divergent dimensionally regulated integrals to
evaluate.

This state of affairs is to be contrasted with the comforting
situation at the NLO level, where general solutions have been
available for long. The approach of choice is that of Catani and
Seymour \cite{Catani:1996vz}, later extended to massive states
\cite{Catani:2002hc} and arbitrary polarizations in real radiation
\cite{Czakon:2009ss}. In fact, we would encourage the non-expert reader
to consult the original paper \cite{Catani:1996vz}, since it gives a
thorough discussion of all aspects of the problem. The present
letter assumes such knowledge.

The main features of the Catani-Seymour subtraction scheme are the
smooth interpolation of the subtraction terms between soft and
collinear limits and independence from the phase space parameterization
achieved by a remapping of phase space points onto the reduced phase
space with one parton less. There is another scheme at NLO derived by
Kunszt, Frixione and Signer (FKS) \cite{Frixione:1995ms}, which is
vastly different on the conceptual side. Here, the phase space is
first decomposed into sectors (originally with the help of the jet
function; an independent decomposition has been proposed in
\cite{Nagy:1996bz}), and then parameterized with energy and angle
variables for easy extraction of the subtraction terms. Precisely
these ideas will turn out to be crucial for the scheme that we shall
derive.

Many approaches have been proposed at NNLO. The most successful ones
are Sector Decomposition \cite{Binoth:2000ps,Anastasiou:2003gr,
  Binoth:2004jv} and Antenna Subtraction
\cite{GehrmannDeRidder:2005cm}. Sector decomposition is conceptually
entirely different from the NLO methods cited above. The idea is to
derive a Laurent expansion for the given amplitude, by first
ingeniously parameterizing the complete phase space, mapping it onto
the unit hypercube, and then dividing it into simplexes, in which
singularities are factorized. The parameterization is adapted to
different singularity structures for different diagram classes. A
detailed description on the particular example of Higgs boson
production can be found in \cite{Anastasiou:2005qj}. The main drawback
is that one has to repeat everything for a new problem, which is
relatively easy, if it involves the same number of particles in the
final state with the same mass distribution, but is expected to be a
major effort otherwise. Antenna subtraction, on the other hand, uses
complete matrix elements of simpler processes as building blocks for
the subtraction. In this way, the integration over the unresolved
particle phase space can be made largely with multi-loop methods. It
is this latter part that involves most work, but the result is general
and can be applied to other processes without modification. The
drawback of this approach is the efficiency loss due to the fact that
azimuthal correlations characteristic of collinear limits are not
taken into account, as the simplified matrix elements correspond to
unpolarized scattering. There are also other methods, which are either
specific to a class of problems, such as that of \cite{Catani:2007vq},
which solves the problem for the production of colorless states, or still
require the integration of the subtraction terms, as for example
in \cite{Weinzierl:2003fx} and \cite{Somogyi:2005xz}.

The purpose of this letter is to present a new approach, which should
provide a method both general and simple to derive. To this end, we
need to specify, which problems are essentially difficult and which
are not. Unlike at NLO, there are in fact two different problems
involving real radiation at NNLO. One involves the emission of one
additional parton (in comparison to leading order) out of virtual
diagrams. Usually called single-real radiation, since only one parton
can become unresolved, it can be treated by any of the NLO methods. Of
course, the subtraction terms require a slight modification. In fact,
we believe that the FKS approach will provide the result with the
least effort. By this argument, we shall ignore single-real radiation
and concentrate on double-real radiation. The latter problem involves
two unresolved partons and only tree-level matrix elements. The method
presented will not depend on the nature of the initial state. Let us,
however, consider the more difficult case of hadronic collisions. Due
to the factorization theorem, the result is a convolution of partonic
cross sections with Parton Distribution Functions (PDFs). As long as
the partonic cross section is an ordinary function, this convolution
can be considered independently (see Section~\ref{sec:phase}). Indeed,
in actual Monte Carlo generators, one first generates the fractions of
hadron momenta to be assigned to the partons and then works in the
center-of-mass frame of the partons, multiplying every event by the
PDF weight only at the end. This is the point of view that we shall
adopt as well. To summarize: we will consider the derivation of the
Laurent expansion of the double-real radiation contributions for fixed
initial parton energies.

The main concept of our approach is to mix some ideas of the FKS NLO
subtraction scheme with those of Sector Decomposition. We will
decompose the phase space in two stages. At the first stage, we will
divide the problem into triple- and double-collinear sectors. Then,
using an energy-angle parameterization of the phase space, we will
perform an ordinary sector decomposition mimicking the physical
singular limits. The crucial novelty is that we will show how to
obtain general subtraction terms from the last sector
decomposition. Here, we will use the knowledge of NNLO singular
behavior of QCD amplitudes as studied in
\cite{Berends:1988zn,Campbell:1997hg,Catani:1998nv} and summarized in
\cite{Catani:1999ss}. While we will not give  explicit expressions for
the subtraction terms, a task impossible in a letter due to the
multitude of cases, it is easy to rederive them following the
description.

In the next section, we will derive the scheme on the example of
massive particle production. The reason for this restriction on the
final state is that this letter will be followed by a companion
publication with process specific information and numerical results
for our first application: top quark pair production. A subsequent
section will, however, present the generalization to arbitrary final
states.

%%%%%%%%%%%%%%%%%%%%%%%%%%%%%%%%%%%%%%%%%%%%%%%%%%%%%%%%%%%%%%%%%%%%%%%%%%%%%%%%

\section{Massive final states at leading order}

%%%%%%%%%%%%%%%%%%%%%%%%%%%%%%%%%%%%%%%%%%%%%%%%%%%%%%%%%%%%%%%%%%%%%%%%%%%%%%%%

\subsection{Phase space}
\label{sec:phase}

Let us assume that there are only two massless partons in the final
state, the other final state particles being massive. In case there is
only one massive final state, the presence of soft singularities leads
to a cross section, which is a distribution in the partonic
center-of-mass energy squared, $s$. We will not take this possibility
into account, since it has already been extensively studied for all
processes of phenomenological interest and is a special case involving
additional complications. The cross section will, therefore, be an
ordinary function of $s$. The considered process corresponds to the
following kinematical configuration
\begin{equation}
p_1+p_2 \rightarrow k_1+k_2+q_1+\dots+q_n \; ,
\end{equation}
with
\begin{equation}
s = (p_1+p_2)^2 \; , \;\;\;\; p_1^2=p_2^2=k_1^2=k_2^2 = 0 \; ,
\;\;\;\; q_i^2 = m_i^2 \neq 0 \; , \;\;\;\; i=1,\dots,n \; , \;\;\;\;
n \geq 2 \; ,
\end{equation}
where $p_1,p_2,k_1,k_2,q_1,...,q_n$ are $d$-dimensional momentum vectors.
The $d$-dimensional phase space can be written as
\begin{equation}
\label{eq:ps1}
\int d \Phi_{n+2} = \int \frac{d^{d-1} k_1}{(2\pi)^{d-1} 2k_1^0}
\frac{d^{d-1} k_2}{(2\pi)^{d-1} 2k_2^0}\prod_{i=1}^n 
\frac{d^{d-1} q_i}{(2\pi)^{d-1} 2q_i^0}  
(2\pi)^d \delta^{(d)}(k_1+k_2+q_1+\dots+q_n-p_1-p_2) \; .
\end{equation}
The above definition suggests a factorization of the phase space for
this problem into the three-particle production phase space of the two
massless partons together with an object with invariant mass $Q^2$,
and a decay phase space of the composite with momentum $Q$ into the
massive particles. Such a factorization is motivated by the fact that
most divergences are due to the vanishing of invariants involving only
massless states momenta. The divergences not belonging to this class
are soft and involve the massive states momenta. In this case, the
inverse propagators responsible for the singularities vanish
proportionally to a linear combination of the energy components of
$k_1$ and $k_2$. This will force us to use these energy components as
part of the phase space parameterization. At this point the phase space
can be written as
\begin{eqnarray}
\int d \Phi_{n+2} &=&  \int 
\frac{d Q^2}{2\pi} \nonumber \\ &\times&
\int \frac{d^{d-1} k_1}{(2\pi)^{d-1} 2k_1^0}
\frac{d^{d-1} k_2}{(2\pi)^{d-1} 2k_2^0}
\frac{d^{d-1} Q}{(2\pi)^{d-1} 2Q^0}
(2\pi)^d \delta^{(d)}(k_1+k_2+Q-p_1-p_2) \nonumber \\ &\times&
\int \prod_{i=1}^n 
\frac{d^{d-1} q_i}{(2\pi)^{d-1} 2q_i^0}  
(2\pi)^d \delta^{(d)}(q_1+\dots+q_n-Q) \; .
\end{eqnarray}
The integrations over $Q$ can be performed by exploiting the
$\delta$-function, which leaves
\begin{eqnarray}
\label{eq:phase0}
\int d \Phi_{n+2} &=& \int \frac{d^{d-1} k_1}{(2\pi)^{d-1} 2k_1^0}
\frac{d^{d-1} k_2}{(2\pi)^{d-1} 2k_2^0}
\nonumber \\ &\times&
\int \prod_{i=1}^n 
\frac{d^{d-1} q_i}{(2\pi)^{d-1} 2q_i^0}  
(2\pi)^d \delta^{(d)}(q_1+\dots+q_n-Q) \nonumber \\ 
&\equiv& \int d \Phi_3 \int d \Phi_n(Q) \; ,
\end{eqnarray}
where
\begin{equation}
\label{eq:mom}
Q = p_1+p_2-k_1-k_2 \; .
\end{equation}
The same result could have been obtained directly (and trivially) from
the original expression given in Eq.~(\ref{eq:ps1}), but we wish to
keep the interpretation of the integral as described above. Notice
that the parameterization of $d \Phi_n(Q)$ will be of no further
concern to us, apart from the fact that it is a continuous function of
$Q$. Let us mention, however, that a particularly suitable approach is
to define $d \Phi_n(Q)$ in the center-of-mass frame of $Q$. In this
case, suitable integration parameters can be chosen to allow the
integration over $d\Phi_n(Q)$ to be the first in
Eq.~(\ref{eq:phase0}).

At this point, we can already derive the boundaries of the
three-particle phase space. From here on, we will work in the
center-of-mass system of the colliding particles. Momentum
conservation from Eq.~(\ref{eq:mom}) implies
\begin{equation}
2 (p_1+p_2)(k_1+k_2)-2 k_1 k_2 = s-Q^2 \; ,
\end{equation}
which can be rewritten as
\begin{equation}
\label{eq:mom2}
2 \sqrt{s}(k_1^0+k_2^0)-2 k_1^0 k_2^0 (1-\cos\theta_{12}) = s-Q^2 \; ,
\end{equation}
where $\theta_{12}$ is the angle between the directions of
$\vec{k}_1$ and $\vec{k}_2$. Let
us introduce the variables $\Delta, \beta, \eta_3, \xi_1, \xi_2$
through
\begin{eqnarray}
\Delta &\equiv& \frac{s-Q^2}{s-\left(\sum_{i=1}^n m_i\right)^2} \; , \\
\beta &\equiv& \sqrt{1-\frac{\left(\sum_{i=1}^n m_i\right)^2}{s}} \; , \\ 
\eta_3 &\equiv& \frac{1}{2}(1-\cos\theta_{12}) \; , \\
k_{1,2}^0 &\equiv& \frac{\sqrt{s}}{2}\beta^2 \xi_{1,2} \; .
\end{eqnarray}
Since $Q^2 \geq \left(\sum_{i=1}^n m_i\right)^2$ we have $\Delta,
\beta, \eta_3 \in [0,1]$. Eq.~(\ref{eq:mom2}) can now be rewritten as
\begin{equation}
\label{eq:cond}
\xi_1+\xi_2-\beta^2\eta_3\xi_1 \xi_2=\Delta \; .
\end{equation}
This equation shows that each of the variables $\xi_1$, $\xi_2$ and
$\eta_3$ can vanish independently, which means that the soft and
collinear limits are independent for any $s$. This is due to the fact
that $n>1$, which implied $\Delta\in [0,1]$ (otherwise $\Delta =
1$). Of course, the independence of the limits can be understood
intuitively by noticing that they all remove one of the massless
partons. Only the removal of the two massless partons at the same time
(double-soft limit) requires flexibility in the choice of $Q^2$, since
then all of the initial state energy is transmitted to the $Q$ system.

To obtain the upper bounds on the variables, we solve
Eq.~(\ref{eq:cond}) for one of the energies (the expressions are
symmetric)
\begin{equation}
\label{eq:upp}
\xi_{1,2} = \frac{\Delta-\xi_{2,1}}{1-\beta^2\eta_3\xi_{2,1}} \; .
\end{equation}
Independently of the other parton, the maximum is obtained when
$\Delta=1$, {\it i.e.} the composite system $Q$ is at
threshold. Furthermore, the absolute maximum of the energy occurs,
when the other parton has vanishing energy
\begin{equation}
(\xi_{1,2})_{max}|_{\xi_{2,1}=0} = 1 \; .
\end{equation}
This case does not lead to any divergences, since there is no phase
space at threshold for the massive $Q$ system. Here, the relative
angle, $\eta_3$, has no relevance. For a finite energy 
of the second parton, the maximum is obtained at $\eta_3=1$, which
corresponds to the two partons being anti-parallel
\begin{equation}
(\xi_{1,2})_{max} = \frac{1-\xi_{2,1}}{1-\beta^2\xi_{2,1}} \; .
\end{equation}
Notice, however, that we will not use this bound explicitly, since
our parameterizations will first specify the angles and only then the
energies, and thus the upper bound will rather be given by
Eq.~(\ref{eq:upp}) with $\Delta=1$. Finally, let us note that the range
of energy integration is split into $\xi_1>\xi_2$ and $\xi_2>\xi_1$ at
\begin{equation}
\xi_{1,2}|_{\xi_1=\xi_2} = \frac{1-\sqrt{1-\beta^2\eta_3}}{\beta^2\eta_3} \; .
\end{equation}
In fact, we can cover the whole energy integration area by two
integration regions defined as
\begin{equation}
\xi_1 > \xi_2 \; : \;\;\;\; \xi_1 \in [0,1] \; , \;\;\;\; \xi_2 \in
   \left[ 0,\min\left(\xi_1, \;
     \frac{1-\xi_1}{1-\beta^2\eta_3\xi_1}\right)\right] \; ,
\end{equation}
and
\begin{equation}
\xi_2 > \xi_1 \; : \;\;\;\; \xi_2 \in [0,1] \; , \;\;\;\; \xi_1 \in
   \left[ 0,\min\left(\xi_2, \;
     \frac{1-\xi_2}{1-\beta^2\eta_3\xi_2}\right)\right] \; .
\end{equation}
We will define a function parameterizing the upper bound as follows
\begin{equation}
\xi_{max}(\xi) = \min\left( 1, \;
\frac{1}{\xi}\frac{1-\xi}{1-\beta^2\eta_3\xi}\right) \; .
\end{equation}
Notice that unless the integration over energies is split as above,
the parameterization of the phase space cannot be symmetric. Of course,
the decisive argument for the split has to do with singularities, but
we will always keep the symmetry of the expressions, as it mimics the
symmetry of the final state in the most complicated case of gluons.

%%%%%%%%%%%%%%%%%%%%%%%%%%%%%%%%%%%%%%%%%%%%%%%%%%%%%%%%%%%%%%%%%%%%%%%%%%%%%%%%

\subsection{Decomposition according to collinear singularities}
\label{sec:dec}

We will now decompose the phase space taking into account the
collinear singularities only. The soft singularities will be treated
in the next step.

Collinear singularities occur, when one or more of the following
conditions are satisfied: $k_1||p_{1,2}$, $k_2||p_{1,2}$,
$k_1||k_2$. Let us introduce a function $\theta_1(k)$, such that
$\theta_1(k) = 1$ if $k||p_1$, and $\theta_1(k) = 0$ if $k||p_2$. In
analogy, we introduce $\theta_2(k)$, which satisfies the same
requirement after swapping $p_1$ and $p_2$. Both functions are
supposed to fulfill $\theta_1+\theta_2=1$ and vanish in the respective
limits fast enough to regulate collinear divergences. The simplest
construction satisfying these requirements is $\theta_1(k) =
\theta(\vec{k} \cdot \vec{p}_1)$.
We can now introduce a first partition of the phase space
\begin{eqnarray}
\label{eq:part1}
1 &=& (\theta_1(k_1)+\theta_2(k_1))(\theta_1(k_2)+\theta_2(k_2)) \nonumber \\
&=& \theta_1(k_1)\theta_1(k_2) + \theta_2(k_1)\theta_2(k_2) + 
\theta_1(k_1)\theta_2(k_2) + \theta_2(k_1)\theta_1(k_2) \; .
\end{eqnarray}
The first two terms allow for singularities depending only on three of
the available momenta. For example, the first term will generate
singularities when: $k_1||k_2||p_1$, $k_1||p_1$, $k_2||p_1$ or
$k_1||k_2$. This is the most complicated case and we will treat it
first (see next subsection). We will concentrate on the
$\{k_1,k_2,p_1\}$ set, since the other one can be obtained by symmetry.

The third and fourth terms in Eq.~(\ref{eq:part1}) will allow for
singularities, when $k_1$ and $k_2$ will be parallel to different
momenta, which means that we can treat the problem as if it were an
iterated next-to-leading order limit. We must, however, be careful
with the situation, when neither of the momenta is close to its
respective limit, but they both tend to each other. In order to
separate further this singularity, we will introduce a third function
$\theta_3(k_1,k_2)$, such that $\theta_3(k_1,k_2) = 1$, when
$k_1||k_2$, and we will assume that $\theta_3(k_1,k_2)$ cancels all
divergences due to $k_{1,2}$ being parallel to $p_{1,2}$.

The final decomposition of the phase space will be given as follows
\begin{eqnarray}
1 &=& \nonumber \\ &&
\left. \begin{array}{l}
 +\theta_1(k_1)\theta_1(k_2) \\
 +\theta_2(k_1)\theta_2(k_2)
\end{array} \right\}
\mbox{triple-collinear sector}
\nonumber \\ &&
\left. \begin{array}{l}
 +\theta_1(k_1)\theta_2(k_2)(1-\theta_3(k_1,k_2)) \\
 +\theta_2(k_1)\theta_1(k_2)(1-\theta_3(k_1,k_2))
\end{array} \right\}
\mbox{double-collinear sector} \nonumber \\ &&
\left. \begin{array}{l}
 +(\theta_1(k_1)\theta_2(k_2)+\theta_2(k_1)\theta_1(k_2))\theta_3(k_1,k_2)
\end{array} \right\}
\mbox{single-collinear sector} \; .
\end{eqnarray}
The first two sectors call for different parameterizations of the phase
space adapted to the singular kinematical configurations. On the other
hand, the third one can be treated along with the first sector, since
the latter allows for the same single-collinear divergence. Within the
first two sectors, the parameterizations will be related by symmetry
with respect to the exchange of $p_1 \leftrightarrow p_2$. 

%%%%%%%%%%%%%%%%%%%%%%%%%%%%%%%%%%%%%%%%%%%%%%%%%%%%%%%%%%%%%%%%%%%%%%%%%%%%%%%%

\subsection{Triple-collinear sector}
\label{sec:triple}

\begin{figure}[ht]
\begin{center}
\includegraphics[width=.40\textwidth]{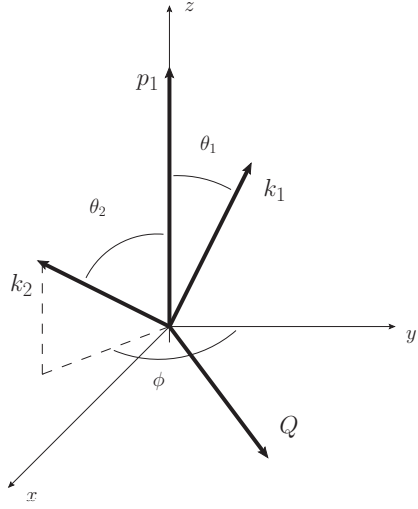}
\end{center}
\caption{Parameterization of the momentum vectors.}
\label{fig:vec}
\end{figure}

We now turn to the case, where the singularities are due to the three
momenta $k_1$, $k_2$ and $p_1$. This situation is depicted in
Fig.~\ref{fig:vec}. We will use the recursive $d$-dimensional definition
of the momentum vectors
\begin{equation}
\vec{n}^{(d)} = (\sin\theta
\; \vec{n}^{(d-1)},\cos\theta) \; .
\end{equation}
Using rotational invariance, we define our momenta as follows
\begin{eqnarray}
\label{eq:moms}
p_1^\mu &=& \frac{\sqrt{s}}{2} (1,\vec{0}^{(d-2)},1)
\; ,  \nonumber \\
k_1^\mu &=& \frac{\sqrt{s}}{2} \beta^2 \xi_1
(1,\vec{0}^{(d-3)},\sin\theta_1,\cos\theta_1) \; ,
 \nonumber \\
k_2^\mu &=& \frac{\sqrt{s}}{2} \beta^2 \xi_2
(1,\vec{0}^{(d-4)},\sin\phi\sin\theta_2,\cos\phi\sin\theta_2,\cos\theta_2)
\; .
\end{eqnarray}
Notice that the sign of $\sin\phi$ is of no relevance for our
considerations. We can, for example, assume that $\sin\phi > 0$ and
only restore the sign using a reflection transformation at the end. On
the other hand, $\theta_{1,2} \in [0,\pi]$ in complete generality. We
will further introduce the notation
\begin{equation}
\eta_{1,2} = \frac{1}{2} (1-\cos\theta_{1,2} ) \; .
\end{equation}
Note that the variable $\eta_3$ defined before is now
\begin{eqnarray}
\eta_3 &=& \frac{1}{2}
(1-\cos\phi\sin\theta_1\sin\theta_2-\cos\theta_1\cos\theta_2)
\nonumber \\ &=& \frac{1}{2}
(1-\cos(\theta_1-\theta_2)+(1-\cos\phi)\sin\theta_1\sin\theta_2)
\; .
\end{eqnarray}
The fact that the relative angle $\theta_{12}$ vanishes only, when
$\theta_1=\theta_2$ and $\phi=0$ is made explicit in the last row. The
three-particle phase space can now be written as follows
\begin{eqnarray}
\int d\Phi_3 &=& \int \frac{d^{d-1} k_1}{(2\pi)^{d-1} 2k_1^0}
\frac{d^{d-1} k_2}{(2\pi)^{d-1} 2k_2^0} \nonumber \\
&=& \frac{1}{8(2\pi)^{5-2\epsilon}\Gamma(1-2\epsilon)} s^{2-2\epsilon}
\beta^{8-8\epsilon} \nonumber \\ &\times&
\int_0^1 d\eta_1 \; (\eta_1(1-\eta_1))^{-\epsilon}
\int_0^1 d\eta_2 \; (\eta_2(1-\eta_2))^{-\epsilon}
\int_{-1}^{1} d\cos\phi \; (1-\cos^2\phi)^{-\frac{1}{2}-\epsilon}
\nonumber \\ &&
\iint d\xi_1 d\xi_2 \; \xi_1^{1-2\epsilon}
\xi_2^{1-2\epsilon} \; ,
\end{eqnarray}
where the energy integration range has been described previously, and
$d=4-2\epsilon$.

As far as the amplitude is concerned, the singular propagator
denominators are
\begin{eqnarray}
-(p_1-k_1)^2 &=& s\beta^2 \xi_1 \eta_1 \; , \nonumber \\
-(p_1-k_2)^2 &=& s\beta^2 \xi_2 \eta_2 \; , \nonumber  \\
(k_1+k_2)^2 &=& s\beta^4 \xi_1 \xi_2 \eta_3 \; , \nonumber \\
-(p_1-k_1-k_2)^2 &=& s\beta^2 (\xi_1\eta_1 + \xi_2\eta_2 
-\beta^2\xi_1\xi_2\eta_3) \; ,
\label{eq:invariants}
\end{eqnarray}
where the signs have been chosen such that the expressions are all
positive definite. We have omitted the propagators of the massive
particles, since they have identical structure, but the analogues of
the $\eta$ variables can never vanish. The first two structures are
already fully factorized. Problems arise only in the third and fourth
cases.

Let us focus on $\eta_3$. We already noted before that the presence of
a singularity requires that there be $\eta_1=\eta_2$ (which is
equivalent to $\theta_1=\theta_2$) and $\phi=0$. This is an example of
a line singularity, since in the three-dimensional space spanned by
$\eta_1,\eta_2$ and $\cos\phi$, the singularity corresponds to a
straight line on the $\cos\phi=1$ plane. On the other hand, a suitable
parameterization would only exploit two variables, in which case the
singular limits would be $\eta_1=0$, $\eta_2=0$ and $\eta_1=\eta_2$,
{\it independently} of the third parameter needed to cover the whole
phase space. In order to obtain such a parameterization, we will
perform a variable change. While the choice is not unique, we will use
a non-linear variable transformation inspired by that used in
\cite{Anastasiou:2005qj} in a similar setting
\begin{equation}
\zeta \equiv \frac{1}{2} \frac{(1 - \cos(\theta_1 - \theta_2))(1 +
  \cos\phi)}{1 - \cos(\theta_1 - \theta_2) + (1 -
  \cos\phi)\sin\theta_1\sin\theta_2} \in [0,1] \; .
\end{equation}
With this variable, the phase space becomes
\begin{eqnarray}
\label{eq:phase2}
\int d\Phi_3 &=& \frac{\pi^{2\epsilon}}{8(2\pi)^5\Gamma(1-2\epsilon)}
s^{2-2\epsilon} \beta^{8-8\epsilon} \int_0^1 d\zeta \;
(\zeta(1-\zeta))^{-\frac{1}{2}-\epsilon} \nonumber \\ &\times&
\iint_0^1 d\eta_1 d\eta_2 \;
(\eta_1(1-\eta_1))^{-\epsilon} (\eta_2(1-\eta_2))^{-\epsilon}
\frac{\eta_3^{1-2\epsilon}}{|\eta_1-\eta_2|^{1-2\epsilon}}
\nonumber \\ && \iint d\xi_1 d\xi_2 \; \xi_1^{1-2\epsilon}
\xi_2^{1-2\epsilon} \; ,
\end{eqnarray}
whereas $\eta_3$ takes the form
\begin{equation}
\eta_3 =
\frac{(\eta_1-\eta_2)^2}{\eta_1+\eta_2-2\eta_1\eta_2-2(1-2\zeta)
\sqrt{\eta_1(1-\eta_1)\eta_2(1-\eta_2)}} \; .
\end{equation}
It can be demonstrated that besides the integrable singularities
in $\zeta$ present explicitly in Eq.~(\ref{eq:phase2}) (which can be
remapped to improve convergence), no other singularities are
introduced into the construction. In other words, $\zeta$ is not
substantial for our discussion.

\begin{figure}[ht]
\begin{center}
\includegraphics[width=.80\textwidth]{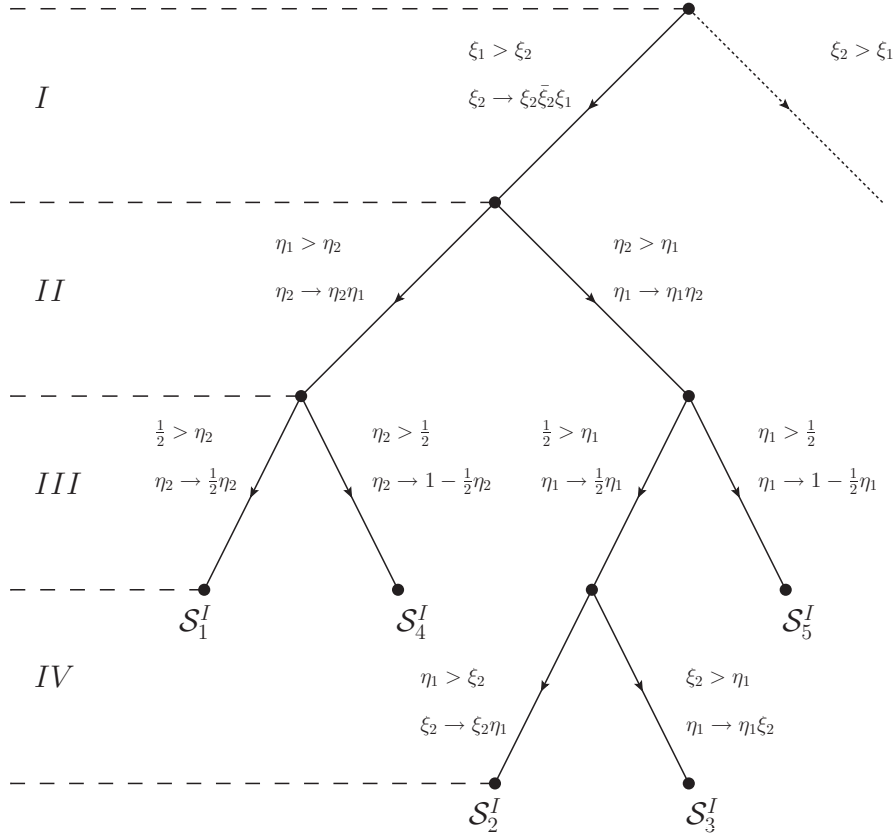}
\end{center}
\caption{Sector decomposition of the phase space in the
  triple-collinear sector. The variable substitutions, which map the
  integration range onto the unit hypercube are
  specified. Furthermore, $\bar \xi_2 = \xi_{max}(\xi_2)$ and the
  second branch starting with the dashed line is symmetric to the
  first.}
\label{fig:dec1}
\end{figure}

At this point, we are ready to introduce a further decomposition of
the phase space that will factorize the invariants given in
Eq.~(\ref{eq:invariants}) and the $\eta_3/|\eta_1-\eta_2|$ factor from
the phase space measure of Eq.~(\ref{eq:phase2}) into a product of the
integration variables and a function, which does not vanish in any of
the singular limits. The decomposition is presented in form of a tree
in Fig.~\ref{fig:dec1}. For a given ordering of the energies of the
massless partons, it contains only five sectors, ${\cal S}^I_1, ...,
{\cal S}^I_5$. In the case of gluons  (most intensive from the
computational point of view), there is no need to consider the other
ordering obtained by symmetry, since the matrix element itself is
symmetric. The levels in the decomposition tree have a clear physical
interpretation:
\begin{itemize}
\item[I)] factorization of the soft singularities;
\item[II, III)] factorization of the collinear singularities;
\item[IV)] factorization of the soft-collinear singularities.
\end{itemize}
Each sector specifies an ordering of the relevant variables, thus
uniquely defining the limits. Collinear singularities require two
levels of decompositions, since besides defining which of the two
partons is allowed to become parallel to $p_1$ first, we need to
single out the possibility that the partons become collinear to each
other first and only then to $p_1$. This is achieved at level III.  An
explicit check proves that in all sectors, the necessary factorization
is reached. At the same time, the inverse propagators of the massive
states responsible for soft singularities are treated at level I.

Let us denote the integration measure obtained from
Eq.~(\ref{eq:phase2}) for a given sector, ${\cal S}$, by $\mu_{\cal
  S}$. We consider the contribution of the phase space integral on
${\cal S}$ to an observable, ${\cal O}$, defined by a measurement
function, customarily called {\it jet function}, $F_J$, acting on the
phase space. We have
\begin{equation}
{\cal O}_{\cal S} = {\cal N} \int_0^1 d\zeta d\eta_1 d\eta_2 d\xi_1
d\xi_2 \; \mu_{\cal S} \; \theta_1(k_1) \theta_1(k_2) \int d\Phi_n(Q)
\; F_J \left| M^{(0)}_{n+2} \right|^2 \; ,
\end{equation}
where ${\cal N}$ is a product of the flux, symmetry factors for the
final state, and spin and color average factors for the initial state.
$M^{(0)}_{n+2}$ is the tree-level amplitude with $n+2$
particles. Summation over final state polarizations has been omitted
in the notation (it is unnecessary, since the formalism is correct for
polarized amplitudes as well). Moreover, as explained in the
Introduction, we have ignored the convolution with the PDFs. By
construction, we can write the following decomposition
\begin{equation}
\mu_{\cal S} \left| M^{(0)}_{n+2} \right|^2 = \sum_{a_i \geq 0}
\frac{1}{\eta_1^{a_1-b_1\epsilon}} \frac{1}{\eta_2^{a_2-b_2\epsilon}}
\frac{1}{\xi_1^{a_3-b_3\epsilon}} \frac{1}{\xi_2^{a_4-b_4\epsilon}}
{\cal M}_{a_1,...,a_4}\; ,
\end{equation}
where both $a_i$ and $b_i$ are integers, $b_i$ being defined by
$\mu_{\cal S}$ alone. To make the decomposition unique, we require first
that ${\cal M}_{a_1,...,a_4}$ be regular in the limit of any of the
$\eta_{1,2},\xi_{1,2}$ variables vanishing. Furthermore, if a given
${\cal M}_{a_1,...,a_4}$ is divided by one of the four variables, {\it
  i.e.} $a_i > 0$ for some $i$, it is not allowed to vanish, when this
variable tends to zero. We would now like to obtain a Laurent
expansion in $\epsilon$ for ${\cal O}_{\cal S}$. The degree of the
singularities given by the $a_i$ is crucial to determine the simplest
possible construction. Usual general arguments on the IR structure of
QCD amplitudes should convince us that $a_i \leq 1$, {\it i.e.} there
are only logarithmic singularities. Thanks to sector decomposition,
this statement can be verified explicitly for any amplitude, as we
will shortly see. Let us introduce
\begin{equation}
{\cal M}_{\cal S} = \eta_1^{1-b_1\epsilon}\eta_2^{1-b_2\epsilon}
\xi_1^{1-b_3\epsilon} \xi_2^{1-b_4\epsilon} \times \mu_{\cal S} \left|
M^{(0)}_{n+2} \right |^2 \; .
\end{equation}
With this definition, the observable becomes
\begin{equation}
\label{eq:main}
{\cal O}_{\cal S} = {\cal N} \int_0^1 d\zeta d\eta_1 d\eta_2 d\xi_1
d\xi_2 \; \theta_1(k_1) \theta_1(k_2) \;
\frac{1}{\eta_1^{1-b_1\epsilon}} \frac{1}{\eta_2^{1-b_2\epsilon}}
\frac{1}{\xi_1^{1-b_3\epsilon}} \frac{1}{\xi_2^{1-b_4\epsilon}} \;
\int d\Phi_n(Q) \; F_J \; {\cal M}_{\cal S} \; ,
\end{equation}
and the Laurent expansion is obtained by means of the replacement
\begin{equation}
\label{eq:sub1}
\frac{1}{\lambda^{1-b \epsilon}} = \frac{1}{b}\frac{\delta(\lambda)}{\epsilon}
+ \sum_{n=0}^\infty \frac{(b \epsilon)^n}{n!} \left[ \frac{\ln^n
    (\lambda)}{\lambda} \right]_+ \; ,
\end{equation}
where $\lambda=\eta_{1,2},\xi_{1,2}$, and the ``+''-distribution is
\begin{equation}
\label{eq:sub2}
\int_0^1 d\lambda \; \left[ \frac{\ln^n (\lambda)}{\lambda} \right]_+
f(\lambda) = \int_0^1 \frac{\ln^n (\lambda)}{\lambda} (f(\lambda) -
f(0)) \; .
\end{equation}
The $f(0)$ term in Eq.~(\ref{eq:sub2}) is to be viewed as the
subtraction term regularizing the amplitude squared, whereas the term
proportional to $\delta(\lambda)$ in Eq.~(\ref{eq:sub1}) is the
integrated subtraction term. The integrated subtraction terms are
integrated over the remaining kinematical variables  alongside the
subtracted amplitude, and no attempt is made at analytic
results. After evaluating the integral over the $\delta$-function, we
can even restore the integration sign, since $\int_0^1 d\lambda =
1$. Notice finally, that the integrated subtraction terms need their
own subtraction, which is generated by the same expression. The result
is obtained by systematically expanding the expression
Eq.~(\ref{eq:main}) (after the substitutions), now a product  of
polynomials in $\epsilon$ (we keep only four terms in
Eq.~(\ref{eq:sub1})), in the latter variable down to the finite
part. This is correct, since the generated integrals are convergent by
construction. To  some extent, viewing this approach as subtraction
terms and integrated subtraction terms is inappropriate, but we wish
to keep the analogy to the traditional approach. 

Up to now, we have worked with an abstract amplitude, which would,
however, have to be specified once a definite process would be
analyzed. At this point we will make a crucial step in the
construction, namely the transition to process independent
subtraction. Thanks to the decomposition in terms of physical
singularities, each subtraction term generated above and corresponding
to one or more of the variables vanishing can be obtained in full
generality by means of the known limiting behavior of QCD amplitudes
as summarized in \cite{Catani:1999ss}. Indeed, ${\cal M}_{\cal S}$
will not vanish in the limit, if and only if there is a singularity of
the amplitude. The value at this point is obtained at the singularity
and the factorization into the tree-level amplitude squared with
partons removed times an eikonal factor or splitting function (divided
by the singular invariant) applies. Thus the subtraction terms will be
given by process specific amplitudes squared (possibly color and/or
spin correlated) taken in their reduced kinematics, multiplied by the
decomposed product of the measure $\mu_{\cal S}$ and eikonal factor or
splitting function (divided by the singular invariant). Notice that
the reduced kinematics is obtained automatically due to our
construction of the phase space (the concept of momentum mapping used
in most subtraction schemes is absent here). Moreover, we do not need
anymore to think of amplitudes, but only of the eikonal factors and
splitting functions. Using their form from \cite{Catani:1999ss}, it is
possible to demonstrate explicitly that ${\cal M}_{\cal S}$ is indeed
always finite, when one or more variables tend to zero. This proves
that divergences are only logarithmic. In summary, the subtraction
terms can be determined once and for all. Due to the multitude of
cases, the full set is substantially larger than in the
next-to-leading order case, but can be derived readily with the
information provided above and in \cite{Catani:1999ss} by means of
simple substitutions (no integration is involved, just simple
algebra). To be specific, one should proceed in two steps:
\begin{itemize}
\item[1)] Substitute Eqs.~(\ref{eq:sub1}) and (\ref{eq:sub2}) into
  Eq.~(\ref{eq:main}), while expanding the ``+''-distributions. The result
  contains 16 different ${\cal M}_{\cal S}$ objects with vanishing
  arguments for $\eta_{1,2},\xi_{1,2}$ in all possible
  combinations. The term with all variables different from zero is the
  full amplitude squared multiplied by the integration measure.
\item[2)] For a given combination of zero arguments of ${\cal M}_{\cal
  S}$, identify the physical limit and take the appropriate
  factorization formula from \cite{Catani:1999ss}. Subsequently,
  calculate the limit using the explicit form of the eikonal factor or
  splitting function. The reduced tree-level amplitude from the
  factorization formula remains unspecified.
\end{itemize}
Finally, let us stress that the subtraction terms are local by
construction, a crucial feature to assure efficiency. The transverse
momentum vectors inducing spin correlations and defining the collinear
limits can be derived from the momentum parameterization of 
Eq.~(\ref{eq:moms}), and read in the single-collinear limits
\begin{eqnarray}
k_{\perp 1}^\mu &=& (0,\vec{0}^{(d-4)},0,1,0) \; , \nonumber
\\ 
k_{\perp 2}^\mu &=&
(0,\vec{0}^{(d-4)},2\sqrt{\zeta(1-\zeta)},2\zeta-1,0)
\; , \nonumber
\\ 
k_{\perp 3}^\mu &=&
(0,\vec{0}^{(d-4)},\sqrt{1-\zeta},\sigma(1-2\eta)\sqrt{\zeta},
-2\sigma\sqrt{\eta(1-\eta)\zeta}) \; ,
\end{eqnarray}
where $k_{\perp 1,2}$ parameterize the transverse direction of
$k_{1,2}$ with respect to $p_1$, whereas $k_{\perp 3}$ parameterizes
the transverse direction of $k_2$ with respect to $k_1$ at
$\eta_1=\eta_2=\eta$ (original variables before sector decomposition),
with $\sigma=(\eta_2-\eta_1)/|\eta_2-\eta_1|$, and we have assumed
$\sin\phi > 0$. In the triple-collinear limit defined by
$\eta_1,\eta_2 \rightarrow 0$, $\eta_1/\eta_2 = $const., the normalization of
the transverse vectors is not arbitrary. Writing
\begin{eqnarray}
k_1^\mu &=& \beta^2 \xi_1 p_1^\mu+\lambda k_{\perp 1}^\mu + {\cal
  O}(\lambda^2) \; , \\ \nonumber
k_2^\mu &=& \beta^2 \xi_2 p_1^\mu+\lambda k_{\perp 2}^\mu + {\cal
  O}(\lambda^2) \; ,
\end{eqnarray}
we can keep $k_{\perp 1}^\mu$ as above, if $\lambda = \frac{\sqrt{s}}{2} \beta^2
\xi_1 \times 2\sqrt{\eta_1}$ and
\begin{equation}
k_{\perp 2}^\mu = \frac{\xi_2}{\xi_1}
\frac{\sqrt{\rho}}{(1-\sqrt{\rho})^2 + 4\sqrt{\rho}\zeta} \times
(0,\vec{0}^{(d-4)},-2\sigma(1-\rho)\sqrt{\zeta(1-\zeta)},2(1+\rho)\zeta
-(1-\sqrt{\rho})^2,0) \; ,
\end{equation}
where $\rho = \eta_2/\eta_1$.

%%%%%%%%%%%%%%%%%%%%%%%%%%%%%%%%%%%%%%%%%%%%%%%%%%%%%%%%%%%%%%%%%%%%%%%%%%%%%%%%

\subsection{Double-collinear sector}
\label{sec:double}

\begin{figure}[ht]
\begin{center}
\includegraphics[width=.80\textwidth]{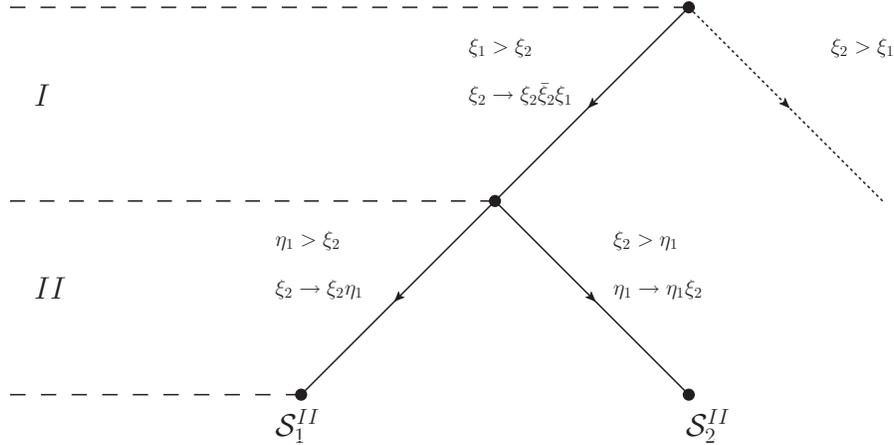}
\end{center}
\caption{Sector decomposition of the phase space in the
  double-collinear sector. The notation is as in Fig.~\ref{fig:dec1}.}
\label{fig:dec2}
\end{figure}

The double-collinear sector can be treated with exactly the same
techniques as the triple-collinear sector. For example, the case where
the phase space is restricted by
$\theta_1(k_1)\theta_2(k_2)(1-\theta_3(k_1,k_2))$, can be obtained
with the previous formulae after the replacement $\eta_2 \rightarrow
1-\eta_2$. Of course, there are no singularities depending on the
value of $\eta_3$, which means that one could avoid the non-linear
variable change. Moreover, the levels II and III from the
decomposition tree of Fig.~\ref{fig:dec1} can be safely removed. In the
end there are only two sectors, as given in Fig.~\ref{fig:dec2}.

%%%%%%%%%%%%%%%%%%%%%%%%%%%%%%%%%%%%%%%%%%%%%%%%%%%%%%%%%%%%%%%%%%%%%%%%%%%%%%%%

\section{Extension to arbitrary final states}

It may seem at first that the case of arbitrary final states will be
substantially more involved than the one discussed above. However, the
only purpose of the restriction to massive states at leading order,
was that the system $q_1,...,q_n$ would not cause any collinear or
soft singularities. This may be enforced in the most general case by a
decomposition of the phase space similar to the one described in
Section~\ref{sec:dec}. The only difference is that we need more than a
constraint on the relative angles.

The double-real radiation at next-to-next-to-leading order is a
correction to a leading order process with two massless partons
less. The remaining partons need to be well separated and have
energies bounded from below. For a jet observable, for example, there
may be different ways to enforce this construction. We will assume,
however, that we are given a definite set of initial and final states
(for a jet observable, there would be a sum over all
sub-processes). That is, we need to identify {\it a posteriori} the
leading order processes possible for the given final state. This is
achieved with the help of a {\it jet algorithm} together with the
requirement that the final state described by jets contain at most two
states less than the initial parton configuration.

In order to be able to use the parameterizations and decompositions of
the phase space introduced previously, we need to extend the division
of the phase space described in Section~\ref{sec:dec}.
Let ${\cal I}$ denote the set of initial state massless partons,
and ${\cal F}$ the set of final state massless partons. The total
number of final states is again assumed to be at least four. We
define two types of selector functions acting on the phase  space:
$\theta_{ij,k}$ and $\theta_{ij,kl}$, both non-negative, with $i,j \in
{\cal F}$ and $i \neq j$, and $k,l \in {\cal I} \cup {\cal F}$ and $k
\neq l$, $k \not\in \{i,j\}$, $l \not\in \{i,j\}$. They are required
to satisfy the following conditions with $m \in {\cal F}$ and $n \in
{\cal F} \cup {\cal I}$
\begin{eqnarray}
\label{eq:cond1}
\lim_{\vec{k}_m || \vec{k}_n}
\theta_{ij,k} &=& 0 \; , \;\;\;\; \mbox{if} \; \{m,n\} \not \subset
    \{i,j,k\} \; , \\ \nonumber
\lim_{k^0_m \rightarrow 0} \theta_{ij,k} &=& 0 \; , \;\;\;\; \mbox{if} \; 
  m \not \in \{i,j\} \; \mbox{and} \; m \; \mbox{is a gluon} \; , \\ \nonumber
\lim_{\vec{k}_m || \vec{k}_n}
\theta_{ij,kl} &=& 0 \; , \;\;\;\;  \mbox{if} \; \{m,n\} \neq
    \{i,k\} \; \mbox{and} \; \{m,n\} \neq \{j,l\} \; , \\ \nonumber
\lim_{k^0_m \rightarrow 0} \theta_{ij,kl} &=& 0 \; , \;\;\;\; \mbox{if} \; 
  m \not \in \{i,j\} \; \mbox{and} \; m \; \mbox{is a gluon} \; ,
\end{eqnarray}
and
\begin{equation}
\label{eq:cond2}
1 = 
\sum_{\begin{array}{c} \mbox{\scriptsize pairs} \\ \scriptstyle i,j
    \in {\cal F} \end{array}}
\sum_{\begin{array}{c} \scriptstyle k \in {\cal I} \cup {\cal F}
    \\ \scriptstyle k \not\in \{i,j\} \end{array}}
\Big[ \theta_{ij,k} + \sum_{\begin{array}{c} \scriptstyle l \in {\cal I} \cup {\cal F}
    \\ \scriptstyle l \not\in \{i,j,k\} \end{array}} \theta_{ij,kl}
  \Big] \; .
\end{equation}
The selector functions thus define a partition of the phase
space. Moreover, inclusion of $\theta_{ij,k}$ in the phase space
integral allows to obtain a Laurent expansion using the
triple-collinear sector decomposition described in
Section~\ref{sec:triple}, assuming that $k_i$ and $k_j$ take on the
role of $k_1$ and $k_2$, whereas $k_k$ that of $p_1$. Similarly,
$\theta_{ij,kl}$ allows to obtain a result using the double-collinear
sector decomposition from Section~\ref{sec:double}, with $k_l$
corresponding to $p_2$.

While there is ample freedom in defining the above selector
functions, it is possible to extend the NLO construction from
\cite{Frixione:2007vw} to satisfy the above requirements. Let us
define
\begin{eqnarray}
d_{ij} &=& \left[ \left( \frac{2 E_i}{\sqrt{s}}\right)
\left( \frac{2 E_j}{\sqrt{s}}\right) \right]^\alpha
(1-\cos\theta_{ij})^\beta \; , \\ \nonumber
d_{ijk} &=& \left[ \left( \frac{2 E_i}{\sqrt{s}}\right)
\left( \frac{2 E_j}{\sqrt{s}}\right)
\left( \frac{2 E_k}{\sqrt{s}}\right) \right]^\alpha
\left[ (1-\cos\theta_{ij})(1-\cos\theta_{ik})(1-\cos\theta_{jk})
  \right]^\beta \; ,
\end{eqnarray}
with $\alpha,\beta > 0$ and $\theta_{ij}$ the angle between
$\vec{k}_i$ and $\vec{k}_j$. Moreover, let
\begin{eqnarray}
h_{i,k} &=& \frac{E_k^\gamma}{E_i^\gamma+E_k^\gamma} \; , \\ \nonumber
h_{ij,k} &=& \frac{E_k^\gamma}{E_i^\gamma+E_j^\gamma+E_k^\gamma} \; ,
\end{eqnarray}
with $\gamma > 0$ and $k \in {\cal F}$, and $h_{i,k} = h_{ij,k} = 1$ if
$k \in {\cal I}$. It can readily be verified that the functions
\begin{eqnarray}
\theta_{ij,k} &=& \frac{1}{{\cal D}} \frac{h_{ij,k}}{d_{ijk}} \; , \\ \nonumber 
\theta_{ij,kl} &=& \frac{1}{{\cal D}}
\frac{h_{i,k}}{d_{ik}}\frac{h_{j,l}}{d_{jl}} \; ,
\end{eqnarray}
with
\begin{equation}
{\cal D} = \sum_{\begin{array}{c} \mbox{\scriptsize pairs} \\ \scriptstyle i,j
    \in {\cal F} \end{array}}
\sum_{\begin{array}{c} \scriptstyle k \in {\cal I} \cup {\cal F}
    \\ \scriptstyle k \not\in \{i,j\} \end{array}}
\Big[ \frac{h_{ij,k}}{d_{ijk}} + \sum_{\begin{array}{c} \scriptstyle l
      \in {\cal I} \cup {\cal F} \\ \scriptstyle l \not\in
      \{i,j,k\} \end{array}}\frac{h_{i,k}}{d_{ik}}\frac{h_{j,l}}{d_{jl}}
  \Big] \; ,
\end{equation}
satisfy the conditions in Eq.~\ref{eq:cond1} and Eq.~\ref{eq:cond2}.

We are now able to give the worst case scenario for the number of
sectors to consider. At the first stage there are ($n_F = |{\cal F}|$
and $n_I = |{\cal I}|$)
\begin{equation}
\left( \begin{array}{c} n_F \\ 2 \end{array} \right) (n_I + n_F - 2)
\; ,
\end{equation}
triple-collinear sectors and 
\begin{equation}
2 \left( \begin{array}{c} n_F \\ 2 \end{array} \right)
\left( \begin{array}{c} n_I+n_F-2 \\ 2 \end{array} \right)
\; ,
\end{equation}
double-collinear sectors. Counting both stages, the final number is
\begin{equation}
2 \times 5 \times 
\left( \begin{array}{c} n_F \\ 2 \end{array} \right) 
(n_I + n_F - 2)
\; + \;
2 \times 2 \times 2
\left( \begin{array}{c} n_F \\ 2 \end{array} \right) 
\left( \begin{array}{c} n_I+n_F-2 \\ 2 \end{array} \right)  \; .
\end{equation}
%

%%%%%%%%%%%%%%%%%%%%%%%%%%%%%%%%%%%%%%%%%%%%%%%%%%%%%%%%%%%%%%%%%%%%%%%%%%%%%%%%

\section{Concluding remarks}

The subtraction scheme developed in the previous sections guarantees
the possibility to automatically obtain double-real radiation
contributions to any observable at NNLO. It is general, just as much
as the NLO subtraction schemes, in the sense that the relevant
construction is independent of the process.

Notice that we did not consider it important to show that with this
construction observables would be finite after combination with the
remaining contributions (double-virtual and real-virtual). On the one
hand, the correctness of the Laurent expansion obtained with the
present method is proven by the correctness of the intermediate
steps. The cancellation of the divergences is then guaranteed by the
Kinoshita-Lee-Nauenberg and by the factorization theorem. On the other
hand, aiming at a proof of cancellation leads to a substantial
complication of the scheme. For us, the simplicity of the numerical
implementation was of paramount importance. A similar philosophy was
already present in the sector decomposition method for phase spaces as
devised in \cite{Anastasiou:2003gr,Binoth:2004jv}. As far as the
cancellation of the divergences in dimensional regularization is
concerned, a comment is, however, in order. In principle, consistency is
guaranteed by the use of the conventional dimensional regularization
scheme. This is also implied by our use of formulae from
\cite{Catani:1999ss}. While our first applications will follow this
approach, it is certainly advantageous to use mixed schemes, which
would allow to compute all tree level matrix elements in four
dimensions, just as it is often done at NLO. Suitable transition
formulae, similar to those of \cite{Catani:1996pk}, will need to be
derived in the future.

Finally, let us stress that our scheme is constructed in such a way
that the momenta of all  partons are available together with the
integration weight, which allows to obtain arbitrary distributions on
the fly. One may wonder, if the number of sectors will not be a
limiting factor to the practical feasibility of the calculation. This
should not be the case, since one should consider the sectors as the
analogues of the usual phase space channels. In a Monte Carlo
implementation one would first choose a sector at random and then the
configuration within the sector. The only true measure of complexity
is the structure of the subtraction terms. Since they have been
derived here using the physical limits it is expected that their
number is minimal, and thus the subtraction scheme optimal. Of course,
extensions are possible, the most trivial being to add a cut-off on
the subtraction phase space. Only practice can show, how important
such features will turn out to be.

%%%%%%%%%%%%%%%%%%%%%%%%%%%%%%%%%%%%%%%%%%%%%%%%%%%%%%%%%%%%%%%%%%%%%%%%%%%%%%%%

\section*{Acknowledgments}

\noindent 
I would like to thank Ch. Anastasiou and G. Heinrich for inspiring
discussions, and Z.~Tr\'ocs\'anyi for interesting correspondence on
the content of this letter. Finally, I am grateful to E.W.N. Glover
for suggesting the name of the subtraction scheme.

\noindent
This work was supported by the Heisenberg and by the Gottfried 
Wilhelm Leibniz programmes of the Deutsche Forschungsgemeinschaft.

%%%%%%%%%%%%%%%%%%%%%%%%%%%%%%%%%%%%%%%%%%%%%%%%%%%%%%%%%%%%%%%%%%%%%%%%%%%%%%%%


\section*{References}

\begin{thebibliography}{00}

\bibitem{Catani:1996vz}
  S.~Catani and M.~H.~Seymour,
  %``A general algorithm for calculating jet cross sections in NLO QCD,''
  Nucl.\ Phys.\  B {\bf 485} (1997) 291
  [Erratum-ibid.\  B {\bf 510} (1998) 503];
  %[arXiv:hep-ph/9605323].
  %%CITATION = NUPHA,B485,291;%%

\bibitem{Catani:2002hc}
  S.~Catani, S.~Dittmaier, M.~H.~Seymour and Z.~Trocsanyi,
  %``The dipole formalism for next-to-leading order QCD calculations with
  %massive partons,''
  Nucl.\ Phys.\  B {\bf 627} (2002) 189;
  %[arXiv:hep-ph/0201036].
  %%CITATION = NUPHA,B627,189;%%

\bibitem{Czakon:2009ss}
  M.~Czakon, C.~G.~Papadopoulos and M.~Worek,
  %``Polarizing the Dipoles,''
  JHEP {\bf 0908} (2009) 085;
  %[arXiv:0905.0883 [hep-ph]].
  %%CITATION = JHEPA,0908,085;%%

\bibitem{Frixione:1995ms}
  S.~Frixione, Z.~Kunszt and A.~Signer,
  %``Three jet cross-sections to next-to-leading order,''
  Nucl.\ Phys.\  B {\bf 467} (1996) 399;
  %[arXiv:hep-ph/9512328].
  %%CITATION = NUPHA,B467,399;%%

\bibitem{Nagy:1996bz}
  Z.~Nagy and Z.~Trocsanyi,
  %``Calculation of QCD jet cross sections at next-to-leading order,''
  Nucl.\ Phys.\  B {\bf 486} (1997) 189;
  %[arXiv:hep-ph/9610498].
  %%CITATION = NUPHA,B486,189;%%

\bibitem{Binoth:2000ps}
  T.~Binoth and G.~Heinrich,
  %``An automatized algorithm to compute infrared divergent multi-loop
  %integrals,''
  Nucl.\ Phys.\  B {\bf 585} (2000) 741;
  %[arXiv:hep-ph/0004013].
  %%CITATION = NUPHA,B585,741;%%

\bibitem{Anastasiou:2003gr}
  C.~Anastasiou, K.~Melnikov and F.~Petriello,
  %``A new method for real radiation at NNLO,''
  Phys.\ Rev.\  D {\bf 69} (2004) 076010;
  %[arXiv:hep-ph/0311311].
  %%CITATION = PHRVA,D69,076010;%%

\bibitem{Binoth:2004jv}
  T.~Binoth and G.~Heinrich,
  %``Numerical evaluation of phase space integrals by sector decomposition,''
  Nucl.\ Phys.\  B {\bf 693} (2004) 134;
  %[arXiv:hep-ph/0402265].
  %%CITATION = NUPHA,B693,134;%%

\bibitem{GehrmannDeRidder:2005cm}
  A.~Gehrmann-De Ridder, T.~Gehrmann and E.~W.~N.~Glover,
  %``Antenna Subtraction at NNLO,''
  JHEP {\bf 0509} (2005) 056;
  %[arXiv:hep-ph/0505111].
  %%CITATION = JHEPA,0509,056;%%

\bibitem{Anastasiou:2005qj}
  C.~Anastasiou, K.~Melnikov and F.~Petriello,
  %``Fully differential Higgs boson production and the di-photon signal through
  %next-to-next-to-leading order,''
  Nucl.\ Phys.\  B {\bf 724} (2005) 197;
  %[arXiv:hep-ph/0501130].
  %%CITATION = NUPHA,B724,197;%%

\bibitem{Catani:2007vq}
  S.~Catani and M.~Grazzini,
  %``An NNLO subtraction formalism in hadron collisions and its application   to
  %Higgs boson production at the LHC,''
  Phys.\ Rev.\ Lett.\  {\bf 98} (2007) 222002;
  %[arXiv:hep-ph/0703012].
  %%CITATION = PRLTA,98,222002;%%

\bibitem{Weinzierl:2003fx}
  S.~Weinzierl,
  %``Subtraction terms at NNLO,''
  JHEP {\bf 0303} (2003) 062;
  %[arXiv:hep-ph/0302180].
  %%CITATION = JHEPA,0303,062;%%

\bibitem{Somogyi:2005xz}
  G.~Somogyi, Z.~Trocsanyi and V.~Del Duca,
  %``Matching of singly- and doubly-unresolved limits of tree-level QCD  squared
  %matrix elements,''
  JHEP {\bf 0506} (2005) 024;
  %[arXiv:hep-ph/0502226].
  %%CITATION = JHEPA,0506,024;%%

\bibitem{Berends:1988zn}
  F.~A.~Berends and W.~T.~Giele,
  %``Multiple Soft Gluon Radiation in Parton Processes,''
  Nucl.\ Phys.\  B {\bf 313} (1989) 595;
  %%CITATION = NUPHA,B313,595;%%

\bibitem{Campbell:1997hg}
  J.~M.~Campbell and E.~W.~N.~Glover,
  %``Double Unresolved Approximations to Multiparton Scattering Amplitudes,''
  Nucl.\ Phys.\  B {\bf 527} (1998) 264;
  %[arXiv:hep-ph/9710255].
  %%CITATION = NUPHA,B527,264;%%

\bibitem{Catani:1998nv}
  S.~Catani and M.~Grazzini,
  %``Collinear factorization and splitting functions for
  %next-to-next-to-leading order {QCD} calculations,''
  Phys.\ Lett.\  B {\bf 446} (1999) 143;
  %[arXiv:hep-ph/9810389].
  %%CITATION = PHLTA,B446,143;%%

\bibitem{Catani:1999ss}
  S.~Catani and M.~Grazzini,
  %``Infrared factorization of tree level QCD amplitudes at the
  %next-to-next-to-leading order and beyond,''
  Nucl.\ Phys.\  B {\bf 570} (2000) 287;
  %[arXiv:hep-ph/9908523].
  %%CITATION = NUPHA,B570,287;%%

\bibitem{Frixione:2007vw}
  S.~Frixione, P.~Nason and C.~Oleari,
  %``Matching NLO QCD computations with Parton Shower simulations: the POWHEG
  %method,''
  JHEP {\bf 0711} (2007) 070
  [arXiv:0709.2092 [hep-ph]].
  %%CITATION = JHEPA,0711,070;%%

\bibitem{Catani:1996pk}
  S.~Catani, M.~H.~Seymour and Z.~Trocsanyi,
  %``Regularization scheme independence and unitarity in QCD cross  sections,''
  Phys.\ Rev.\  D {\bf 55} (1997) 6819.
  %[arXiv:hep-ph/9610553].
  %%CITATION = PHRVA,D55,6819;%%

\end{thebibliography}
\end{document}